\documentstyle[prb,aps,graphics]{revtex}
\begin{document}
%\draft
\title{On the Role of Exchange Interaction in Magnetic Ordering and Conductivity of 
Manganites}
\author{M.V.Krasinkova \cite{Auth}}
\address{Ioffe Physical Technical Institute, Russian Academy of Sciences, 
St.Petersburg 194021, Russia}
\maketitle
\begin{abstract}
A model of chemical bonding between ions in manganites involving covalent 
one-electron $\sigma$ bonding is suggested. The covalent one-electron $\sigma$ 
bonding gives rise to a strongly correlated state of electrons resulting 
from the exchange interaction between electrons when they are simultaneously 
at cation and anion orbitals. The manifestation of the correlatred state is the
spin and spatial ordering of the electrons resulting in the formation of a spin-
ordered electron lattice. The conductivity of manganites in this model is the 
consequence of displacement of the electron lattice (or its part) from one localization 
site to another and depends on the type of spin ordering of the electrons in 
the electron lattice and the localization energy determined by the 
energy of the one-electron $\sigma$ bond. The model also assumes a strong
polarization of an anion by cations, which facilitates the 3s2p hybridization of the 
anion and transition of one of the pairs of 2p electrons from the singlet state 
into the triplet state. This transition leads to formation of the spin-polarized
electron lattice (electron spins are parallel) and ferromagnetic ordering of 
manganese ions. In the model, the effect of colossal magnetoresistance 
is explained by a change of the conductivity mechanism on application of an 
external magnetic field, i.e., transition from the conductivity mechnism typical 
of an ionic crystal to the conductivity provided by the spin-polarized electron
lattice.
\end{abstract}

\section{INTRODUCTION}
Electric and magnetic properties of doped manganites have received much attention of 
researchers in recent years because these compounds are thought to be promising for 
the use as high-sensitivity detectors of magnetic fields, in magnetic resistive memory, 
readout magnetic heads, and in devices based on the spin-polarized transport. \cite{1} 

Of special interest among doped manganites are the compounds where transition to 
the magnetic ordering is accompanied by a metal-insulator transition. \cite{2} The 
interplay of magnetic and transport properties in manganites has been explained for a 
long time in the framework of the double-exchange theory \cite{3} which assumes 
electron exchange between neighboring $Mn^{3+}$ and $Mn^{4+}$ ions via an 
intervening $O^{2-}$ ion. 

However, experimental data obtained in recent years, such as an increase in $T_c$ under 
an external magnetic field \cite{4}, the effect of colossal magnetoresistance \cite{5},  
coexistence of charge ordering and ferromagnetism \cite{6}, ferromagnetic ordering 
in the compounds $CaCu_3MnO_{12}$ \cite{7} and $Tl_2Mn_2O_7$ \cite{8} where 
there are no $Mn^{3+}$ ions, have thrown doubt on the concept of double exchange as 
a basic mechanism of ferromagnetic ordering and conductivity in manganites \cite{9}.

An important achievement in studies of doped manganites was detection of local 
distortions of the crystal lattice below $T_c$.\cite{10} It was shown that the actual 
crystal lattice strongly differs from the averaged lattice identified by X-ray diffraction. 
These experimental data have led to the conclusion on a close relation between 
magnetic and structural properties. However, the questions of how the local lattice 
distortions  influence the electron structure and magnetic ordering and what is the 
origin of these distortions still remain unresolved. 

The goal of this work was to show that the interplay between magnetic, transport, and 
structural properties can be explained if we take into account a strong covalency of 
bonding between manganese and oxygen ions and represent it as formation of one-
electron $\sigma$ bonds between these ions. This concept of one-electron $\sigma$ 
bonding in manganites was earlier invoked to explain antiferromagnetic ordering. 
\cite{11} In the paper it is shown that this concept can prove useful for explanation of 
ferromagnetic ordering as well. Moreover, this work has a more general purpose - to 
show by using manganites as an example how a strongly correlated state of the 
electrons responsible for magnetic ordering and high conductivity can arise via the 
exchange interaction between the electrons forming one-electron $\sigma$ bonds.

\section{Covalent one-electron bonding}

As for the type of chemical bonding, manganites occupy an intermediate position 
between ionic and covalent crystals. This is evidenced by the difference between 
electronegativities (1.55 for Mn and 3.44 for O according to Pauling \cite{12}) equal 
to 1.9, which corresponds to approximately 40\% of covalency and 60\% of ionicity of 
bonding between Mn and O atoms. A strong covalency in manganites has been 
recently experimentally confirmed by high-resolution Mn $k_\beta$ emission 
spectroscopy. \cite{13} Such covalency must manifest itself first of all in the behavior 
of valence electrons because covalency means sharing of electrons by adjacent ions, 
while ionicity requires that they be localized at anions. The question arises how the 
state of valence electrons in a solid when they are simultaneously localized at anions 
and shared with cations can be imagined. 

For molecules, this type of bonding was regarded as a resonance state of structures 
with a purely covalent and purely ionic bond. \cite{12} However, it is hardly probable 
that this approach is appropriate for solids because it is unlikely that a typical two-
electron $\sigma$ bond considered in the resonance model is possible in an ionic crystal. 
In addition, it is difficult to represent ordering of different types of bonds. Probably, 
it would be more reasonable to treat  such 
bonding as {\em sharing of only one electron} by each cation-anion pair. 

Let us suppose that at high temperatures, ions are mainly coupled ionically, and this
ionic bonding determines the averaged crystal structure of manganites with a high-symmetry 
lattice of ions. As temperature decreases, additional covalent one-electron  $\sigma$ 
bonding between Mn and O ions appears. In this case the crystal lattice undergoes 
distortions, and the crystal symmetry is lowered. Appearance of covalent one-electron 
bonding between ions means a partial substitution of ionic bonding by covalent 
bonding, i.e., the Coulomb interaction energy between ions is partly substituted by the 
covalent bonding energy. As a result, the bonding energy increases, which is 
evidenced by a decrease in the distance between ions which turns to be less than the 
sum of the ionic radii. Let us treat this covalent one-electron bonding as a {\em set of 
localized bonds} between neighboring ions. 

If Mn ions are surrounded by oxygen octahedra, they can use six equivalent hybrid 
$3d^24s4p^3$ orbitals for covalent bonding. Since the energy level of these hybrid 
orbitals is higher than the level of $d_{xy}$, $d_{yz}$, and $d_{xz}$ orbitals, which results 
from both admixing of a higher-energy 4p state and splitting of d levels in the 
octahedral crystal field, it can be supposed that the difference between the energy 
levels of hybrid and unmixed orbitals is sufficiently large to make the transformation 
of $Mn^{3+}$ ion from the high-spin state $(t_{2g}^3 e_g^1)$ into the {\em low-spin state} 
$(t_{2g}^4)$ energetically favorable. As a result, all six hybrid orbitals of $Mn^{3+}$ 
ions will be empty and can be available for covalent bonding. The probability of the 
low-spin state for $Mn^{3+}$ ion in manganites was also discussed in Ref.\cite{14}.  

An $O^{2-}$ ion in the manganite structure is also octahedrally surrounded, two of the 
surrounding cations being Mn ions lying along one of the axes of the octahedron. To 
provide one-electron $\sigma$ bonding with these cations, the $O^{2-}$ ion can use its 2p 
orbital or{\em two hybrid 3s2p orbitals} pointed to Mn ions. The use of the {\em 2p orbital} has 
already been discussed \cite{11}, but the possibility of 3s2p hybridization needs an 
explanation. Due to a large difference between energies of 2p and 3s levels in an 
oxygen {\em atom}, the 3s2p hybridization is thought to be impossible. However, for the 
$O^{2-}$ ion that is in the {\em ionic crystal lattice}, there are some factors that facilitate such 
hybridization. First, the energy level of the 2p state in the $O^{2-}$ ion is higher than 
that in the oxygen atom, which is evidenced by a difference between the ionic oxygen 
radius and its atomic radius (1.26A and ~0.9A, respectively). Second, $O^{2-}$ ion 
located between two manganese ions with a high positive charge and relatively small 
ionic radius will be strongly {\em polarized} by these two cations. The energy of this 
polarization can partly compensate for the energy spent on excitation of two 2p 
electrons into the 3s2p state. At last, the energy spent on hybridization will also be 
compensated by the total gain in energy due to formation of a stronger bonding 
between ions.     

Let us compare covalent one-electron bonds involving 2p and 3s2p orbitals of the 
$O^{2-}$ ions. We designate them as $1e-\sigma$ and $1e-\sigma^\ast$, respectively. Since 
the $1e-\sigma$ bond is formed by a shorter 2p orbital, its energy will be higher than the 
energy of the $1e-\sigma^\ast$ bond, and therefore the $1e-\sigma$ bond is more favorable 
for the crystal stability. However, the $1e-\sigma$ bond can be formed only if  a cation 
and an anion can closely approach each other to provide overlapping of the cation 
orbital with the 2p anion orbital. The possibility of such overlapping depends  on the 
deformabilily of the ionic lattice and also on the filling of the cation electron shell. 
{\em Nonsphericity} of the electron shell of a $Mn^{3+}$ ion in the low-spin state resulting 
from the presence of a $t_{2g}$ electron pair in one of the planes hinders a close 
approach of this cation and anion in this plane because of a strong repulsion between 
the electron pair and the anion. For this reason, in this plane, only overlapping with 
a more extended 3s2p orbital and, hence, formation of only $1e-\sigma^\ast$ bond is possible. 

Another difference between $1e-\sigma$ and $1e-\sigma^\ast$ bonds is the spin state of 
the anion electron pair that forms these bonds. For $1e-\sigma$ bonds, the pair is in the 
{\em singlet} state because it belongs to one p orbital (Pauli exclusion principle); and 
for $1e-\sigma^\ast$ bonds the pair is in the {\em triplet} state because in hybridization 
two electrons of one 2p orbital must pass into two hybrid 3s2p orbitals, and their spins 
must be parallel according to the Hund's rule because these orbitals are orthogonal and share a space. 

\section{Spin ordering of electrons and electron crystallization}

If two one-electron bonds are formed by one anion with two neighboring cations, the 
spin state of the anion electron pair participating in covalent bonding will determine 
the orientation of electron spins in the hybrid orbitals of these cations (Fig.\ 
\ref{fig1}). 
\\

\includegraphics*[0in,0in][7in,3in]{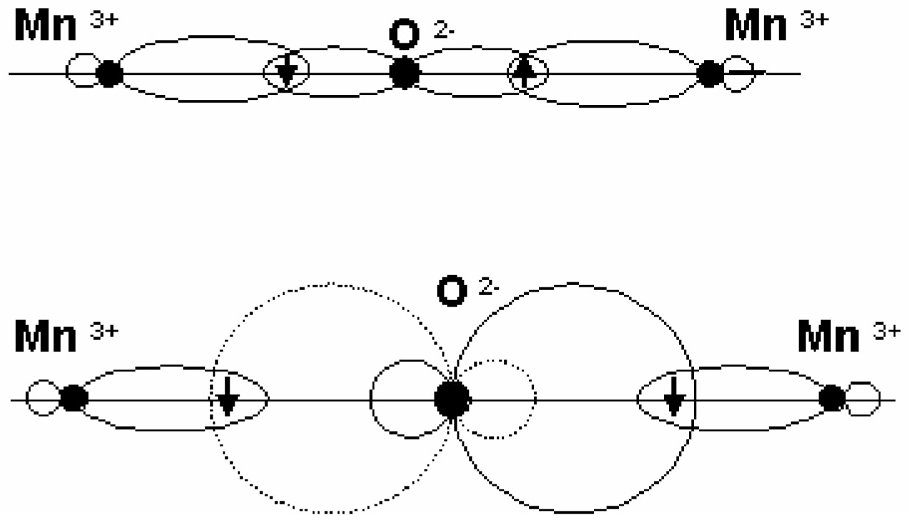}

\begin{figure}
\caption{Formation of covalent one-electron bonding between $Mn^{3+}$ and $O^{2-}$ 
ions: (a) $1e-\sigma$ and (b) $1e-\sigma^\ast$. Empty $d^2sp^3$ hybrid orbitals of two 
$Mn^{3+}$ ions and the p orbital occupied by a singlet electron pair of $O^{2-}$ ion (a) 
and empty $d^2sp^3$ hybrid orbitals of two $Mn^{3+}$ ions and two 3s2p hybrid orbitals 
occupied by a triplet pair of $O^{2-}$ ion (b) are shown.}
\label{fig1}
\end{figure}

Each cation which has six hybrid orbitals can share six electrons - one 
electron from each of six neighboring anions. The spins of all these six electrons must 
be parallel according to the Hund's rule for hybrid $d^2sp^3$ cation orbitals. Therefore, 
if all the anions and all Mn ions are coupled by one-electron $\sigma$ bonds, all the 
electrons taking part in formation of these bonds will be spin-ordered (Fig.\ 
\ref{fig2}). 
\\

\includegraphics*[0,0][6in,1.5in]{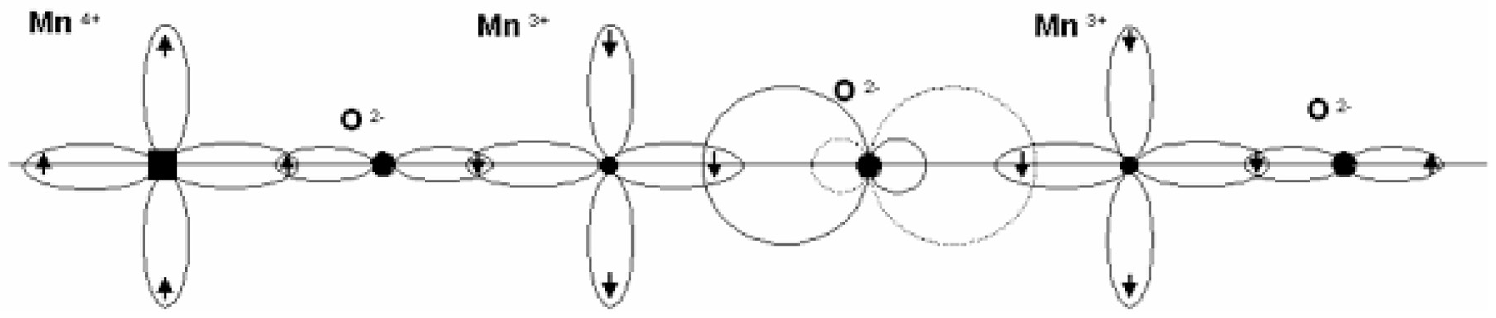}

\begin{figure}
\caption{Electron spin ordering in hybrid orbitals of $Mn^{3+}$ (black circles) and 
$Mn^{4+}$ (a black square) ions at formation of $1e-\sigma$ and $1e-\sigma^\ast$ bonds 
via intervening $O^{2-}$ ions (singlet and triplet pairs). For simplicity, only four of six 
hydrid orbitals for each cation and only three anions are shown.}
\label{fig2}
\end{figure}

Since the Hund's rule reflects the dependence of the electron system 
energy on the mutual spin orientation, and such dependence is equivalent to the 
existence of an additional interaction between electrons which is known as the 
exchange interaction, it can be stated that spin ordering of the electrons participating 
in covalent bonding results from two exchange interactions - between electrons in the 
anion orbital (or two hybrid orbitals) and between electrons in the cation hybrid 
orbitals. {\em In the case of one-electron bonding}, the sign of the exchange interaction in 
the cation hybrid orbitals is {\em always positive}, while the sign of the exchange interaction 
between the electrons in the anion orbitals, as shown above, can be {\em positive} (a triplet 
pair) or {\em negative} (a singlet pair) (Fig.\ \ref{fig2}).     

In the case of positive signs of the exchange interactions between electrons in both the 
anion and cation orbitals, the spins of all the electrons participating in one-electron 
bonding will be parallel, i.e., there will be 100-\% spin polarization of the electrons. In 
the case of different signs of the exchange interactions in cation and anion orbitals, the 
electrons are also spin-ordered, but their spins are antiparallel in the hybrid orbitals of 
neighboring cations. 

The exchange interaction leads to not only spin ordering of electrons, but also spatial 
ordering of electrons relative to each other, which can probably be regarded as the 
electron lattice formation (electron crystallization). On the whole, such ordering 
resembles the Wigner crystal formation \cite{15}, but there is a significant difference. 
The Wigner crystal involves a purely Coulomb repulsion between electrons in a 
uniform field of positive charge. As a result, the Wigner crystal is possible only at low 
temperatures when the Coulomb repulsion energy exceeds the electron thermal 
energy, and at these temperatures the crystal acquires the ability to displace as one 
unit (in the absence of the inhomogeneities that cause pinning). In the case of 
formation of the electron lattice considered in this paper, the repulsion energy is a {\em sum} 
of the purely Coulomb repulsion energy and the repulsion energy due to the exchange 
interaction between electrons with parallel spins in hybrid orbitals. This leads to 
electron crystallization at such high temperatures as $T_c$. Moreover, participation of 
electrons in covalent bonding causes localization of the electron lattice. The energy of 
this localization depends on the one-electron bond energy. Hence, the electrons 
participating in one-electron bonding between ions are in a {\em strongly correlated state} - 
they form a {\em localized spin-ordered electron lattice}. 

It is important to mention here that magnetic ordering of inherent magnetic moments 
of cations resulting from unpaired $t_{2g}$ electrons will be established via a purely 
magnetic interaction between these magnetic moments and the magnetic moment of 
the spin-ordered electron lattice. In the case of its 100-\% spin polarization (spins are 
parallel), magnetic moments of all the cations will be ordered ferromagnetically.  

\section{Conditions for conductivity}

Let us consider now the conductivity in the system of strongly correlated electrons. It 
is likely that, in the presence of strong repulsion between spin-polarized electrons 
(there cannot be two electrons with parallel spins in one orbital), the only possible conductivity 
mechanism is displacement of the electron lattice as one unit, i.e., {\em collective transport 
of all the electrons from one localization site to a neighboring site}. However, such collective displacement is possible if 
transition of any electron to a hybrid orbital of a neighboring cation is not forbidden 
by the spin direction in this orbital and if the localization energy of the electron lattice 
is not too high. The first condition can be satisfied if the electron lattice is spin-
polarized (spins are parallel).  The second condition can be fulfilled by a decrease in 
the extent of overlapping between cation and anion hybrid orbitals, i.e., by an increase 
in the distance between $Mn^{3+} - O^{2-} - Mn ^{3+}$ ions, for instance, due to 
local crystal lattice distortions observed in doped manganites. However, these conditions 
contradict each other because weakening of covalent bonds and their breaking due to 
thermal ion vibrations can lead to a loss of spin ordering between electrons of 
neighboring cations since spin ordering is created and maintained just by the covalent 
bonding. Nevertheless, these two conditions can be satisfied simultaneously if the 
functions of conductivity and maintaining spin ordering are fulfilled by different parts 
of the spin-polarized lattice. This can be realized if a part of the spin-polarized electron 
lattice (spins are parallel) remains three-dimensional and localized, while the second 
part has a lower localization energy and is characterized by a lower dimensionality. 
The first part of the lattice will provide spin ordering, and the second part will provide 
conductivity.  Below we show how such a separation of functions between the parts of 
a spin-polarized electron lattice is achieved in doped manganites.

\section{Magnetic ordering and conductivity in doped and undoped manganites}

Let us now see how magnetic ordering and conductivity of doped and undoped 
manganites can be explained in the framework of the suggested model.

An experimentally observed perovskite cubic lattice of $CaMnO_3$  with a distance 
between ions less than the sum of ionic radii (1.865A and 1.93A, respectively) and 
antiferromagnetic ordering of $Mn^{4+}$ ions below $T_N$ are in good 
agreement with the assumption of six equivalent $1e-\sigma$ bonds formed by each 
$Mn^{4+}$ ion with six surrounding anions via electrons of singlet pairs. The electrons 
participating in covalent bonding and being in hybrid cation orbitals prove to be 
mutually ordered, and their state can also be regarded as an electron lattice. The 
electron lattice in this case consists of electron octahedra surrounding  $Mn^{4+}$ ions, 
the electron spins being parallel in each octahedron and antiparallel with respect to a 
neighboring octahedron below $T_N$. Conductivity is impossible because of different spin 
orientations of electrons in the orbitals of neighboring cations, consistent with 
experimental data. \cite{2} 

As to $LaMnO_3$, the nonsphericity of the electron shell of $Mn^{3+}$ ion, as discussed 
above, results in formation of four $1e-\sigma^\ast$ bonds in the plane where a pair
of $t_{2g}$ electrons is located, which must lead to ferromagnetic ordering between 
neighboring cations. In the direction normal to this plane, two $1e-\sigma$ bonds can 
be formed, and their formation will result in antiferromagnetic ordering between 
neighboring cations. The orientational ordering of 
$Mn^{3+}$ ions in one plane provides ferromagnetic ordering of all $Mn^{3+}$ ions in 
this plane, the magnetic ordering between such planes being antiferromagnetic. 
Precisely this magnetic ordering was observed experimentally below $T_N$. \cite{2} 
The orientational ordering of $Mn^{3+}$ ions in one plane decreases the deformational 
energy of the ionic lattice when bonds of different lengths ($1e-\sigma^\ast$ and $1e-
\sigma$) are formed. Such ordering is energetically favorable. 

The spin-ordered electron lattice in this case consists of electron octahedra 
surrounding $Mn^{3+}$ ions. The electron spins are parallel in all the electron 
octahedra lying in the plane of the orientational ordering of $Mn^{3+}$ ions and are 
antiparallel to the spins of the electron octahedra in the neighboring planes. $LaMnO_3$ 
is an insulator \cite{2}, and this is likely to be due to both a high localization energy 
of the electron lattice in the absence of local lattice distortions and a high 
dimensionality of the spin-polarized lattice (two-dimensional). 

In doped manganites which contain both ions ($Mn^{3+}$ and $Mn^{4+}$), each ion will 
tend to form the bonds similar to those in undoped compounds. As to coupling 
between $Mn^{3+}$ and $Mn^{4+}$ ions via an intervening anion, the $1e-\sigma$ bond is 
likely to be formed between them because it is energetically favorable, and both ions 
are capable of forming such bond. Such bonding leads to antiferromagnetic ordering 
of $Mn^{3+}$ and $Mn^{4+}$ ions. Since a $Mn^{3+}$ ion cannot form all the six $1e-
\sigma$ bonds, this ion will be coupled with the neighboring $Mn^{3+}$ ions by $1e-
\sigma^\ast$ bonds to result in ferromagnetic ordering between these ions. Between 
$Mn^{4+}$ ions, $1e-\sigma$ bonds can be formed if deformation of the ionic lattice (or 
local distortions) is possible, or $1e-\sigma^\ast$ bonds can be formed if such 
deformation is impossible.

In this respect, the $A_{0.67}B_{0.33}MnO_3$ chemical composition of doped manganites 
is unique because it makes possible such charge ordering of $Mn^{3+}$ and $Mn^{4+}$ 
ions in which each $Mn^{4+}$ ion is surrounded by six $Mn^{3+}$ ions, and each 
$Mn^{3+}$ ion is surrounded by {\em three $Mn^{4+}$ ions and three $Mn^{3+}$ ions}. In the 
latter case the cations with equivalent oxidation states are the nearest neighbors (Fig.\ 
\ref{fig3}).  
\\

\includegraphics*[0,0][6in,5.7in]{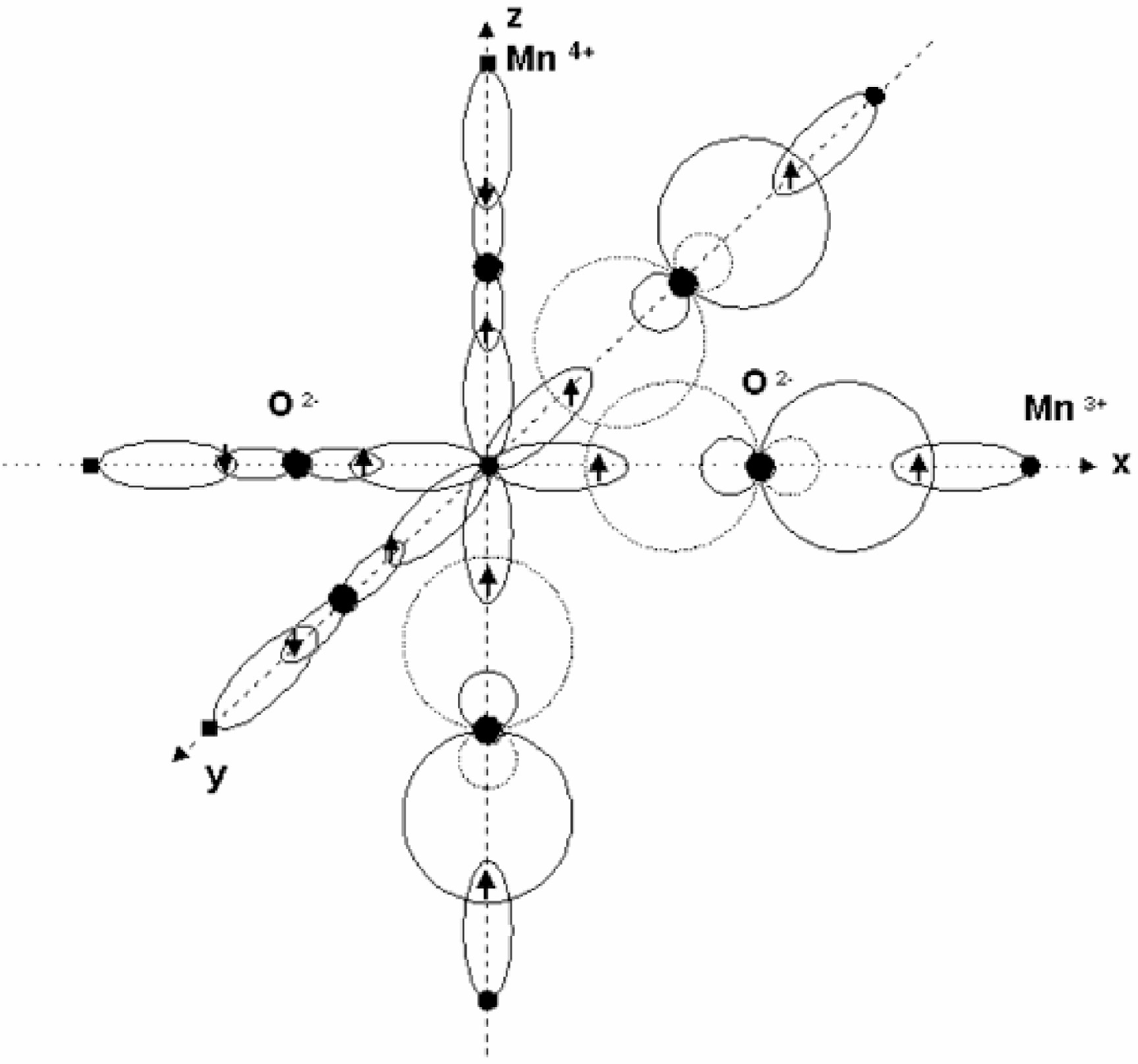}

\begin{figure}
\caption{Coordination polyhedron of a $Mn^{3+}$ ion in $A_{0.67}B_{0.33}MnO_3$. 
Three pairs of $1e-\sigma$ bonds with neighboring $Mn^{4+}$ ions and three pairs of 
$1e-\sigma^\ast$ bonds with neighboring $Mn^{3+}$ ions via an intervening $O^{2-}$
ions are shown.}
\label{fig3}
\end{figure}

Such charge ordering is accompanied by formation of $1e-\sigma^\ast$ 
bonds between all $Mn^{3+}$ ions via intervening $O^{2-}$ ions and $1e-\sigma$ bonds 
between all $Mn^{3+}$ and $Mn^{4+}$ ions. $Mn^{4+}$ ions are not the nearest neighbors 
to each other. In Fig.\ \ref{fig4}, two sublattices formed by $Mn^{3+}$ and $Mn^{4+}$ ions 
separately due to their charge ordering can be seen. The electron lattice involves in 
this case two electron sublattices with antiparallel spins. 
\\

\includegraphics*[0in,0in][6in,5.5in]{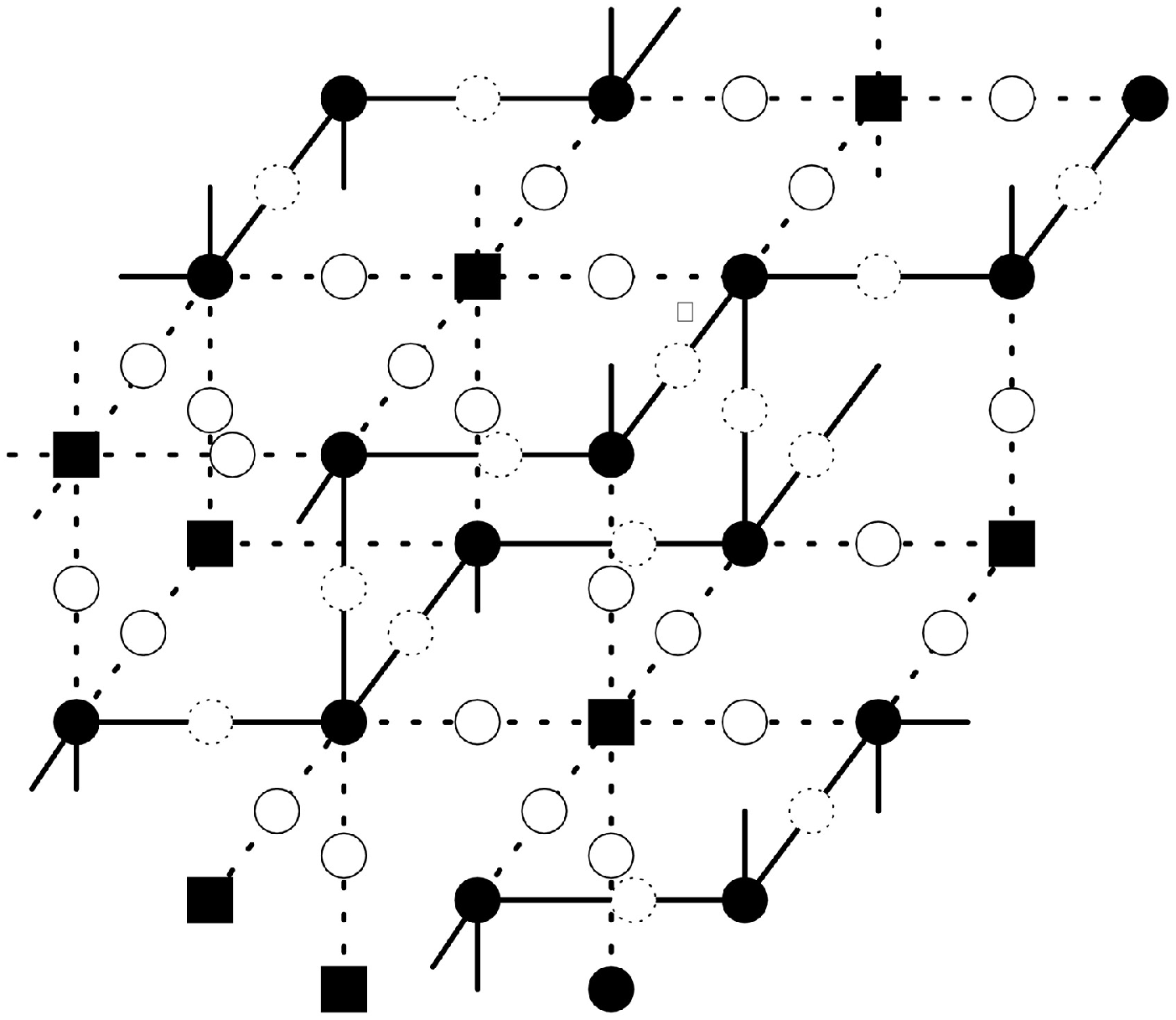}

\begin{figure}
\caption{Ordering of $Mn^{3+}$ ions (black circles), $Mn^{4+}$ ions (black squares), 
$1e-\sigma$ bonds formed by singlet pairs (dashed lines), and $1e-\sigma^\ast$ bonds 
formed by triplet pairs (solid lines) in $A_{0.67}B_{0.33}MnO_3$. Dashed open circles 
show $O^{2-}$ ions in the 3s2p hybrid state. Local distortions are not shown.}
\label{fig4}
\end{figure}

One sublattice consists of 
electron octahedra around $Mn^{4+}$ ions, and the other electron sublattice comprises 
electron octahedra around $Mn^{3+}$ ions. Such classification of the electron 
sublattices is convenient for determining the type of magnetic ordering of intrinsic 
magnetic moments of cations. In this case $Mn^{3+}$ ions themselves and also $Mn^{4+}$ ions 
themselves are ordered ferromagnetically, and antiferromagnetically with respect 
to each other. 

On the other hand, to {\em explain conductivity}, it would be useful to divide the spin-
ordered electron lattice into two sublattices formed by electrons of triplet pairs and  
singlet pairs of anions separately because there is a difference between the 
localization energies of the electrons of these pairs due to the different energies of $1e-
\sigma^\ast$ and $1e-\sigma$ bonds. Such division of the electron lattice allows one to 
separate out the less localized part of the spin-polarized electron lattice (spins are 
parallel). With such choice of electron sublattices, it should be borne in mind that the 
spins of all the electrons in the sublattice of singlet pairs are ordered.  As can be seen 
from Figs.\ \ref{fig3} and\ \ref{fig4}, the electron sublattice of triplet pairs is 
a part of the spin-polarized lattice consisting of electron octahedra around $Mn^{3+}$
and turns out to be quasi-one-dimensional in each $MnO_2$ plane. It is arranged between 
the nearest $Mn^{3+}$ ions in the form of stripes parallel to each other. These 
stripes are separated from each other in each $MnO_2$ plane by the stripes of the 
electron sublattice of singlet pairs. Since formation of short $1e-\sigma$ bonds between 
$Mn^{3+}$ and $Mn^{4+}$ ions results in an increase in the distance between $Mn^{3+}$ 
ions (local distortions, see Fig.\ \ref{fig5}), which is accompanied by the decrease in 
overlapping and hence in the energy of the $1e-\sigma^\ast$ bond, it can be supposed 
that in this manganite composition the electron sublattice of triplet pairs becomes less 
localized and more capable of displacing from one localization site to a neighboring 
one compared with $LaMnO_3$. 
\\

\includegraphics*[0in,0in][6in,4.5in]{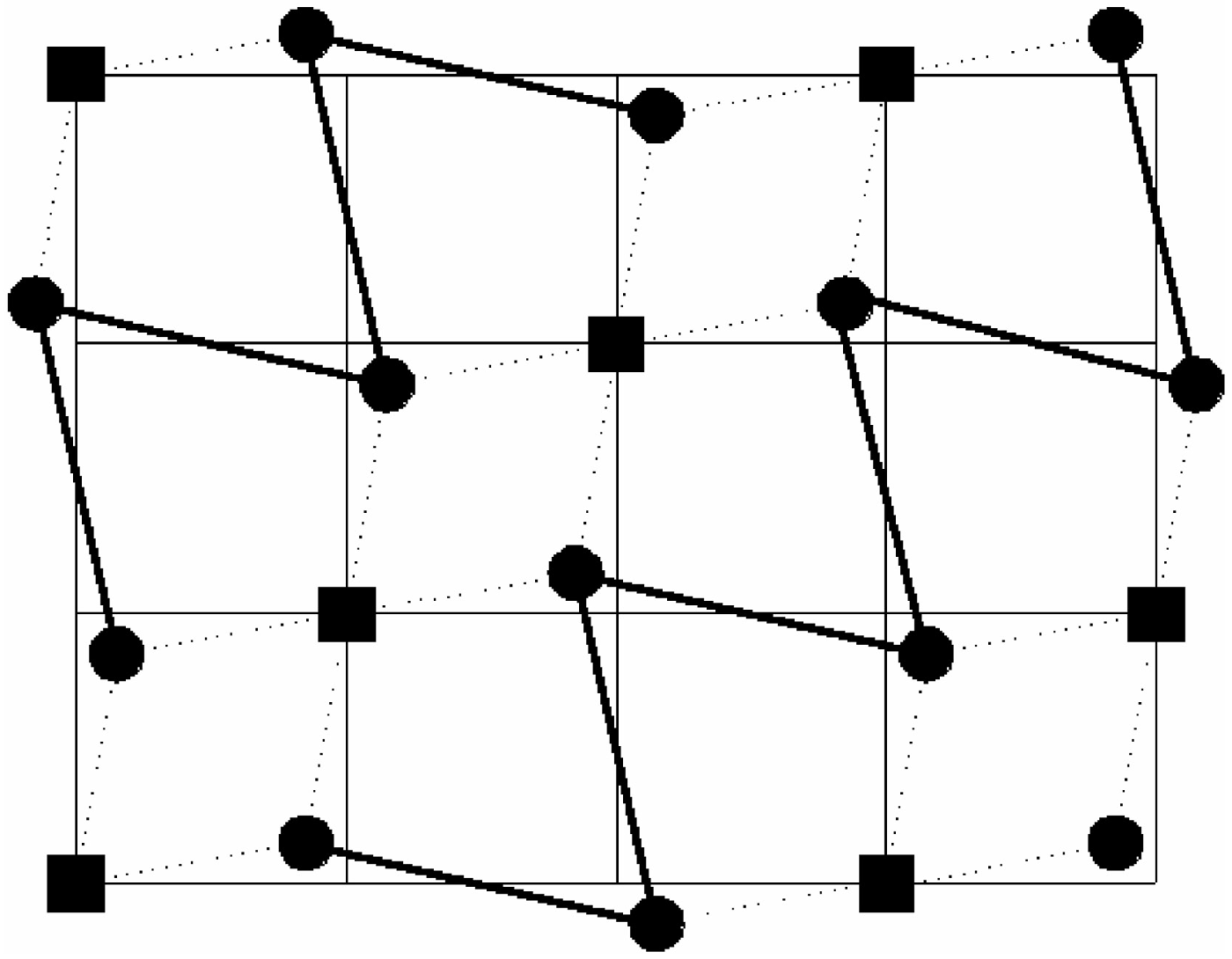}

\begin{figure}
\caption{Projection of a distorted crystal lattice onto a $MO_2$ plane.  
The $1e-\sigma$ bonds and $1e-\sigma^\ast$ bonds formed by siglet and triplet
electron pairs are shown. The $O^{2-}$ ions forming the bonds are not shown. The 
designations are the same as in Fig.\ \ref{fig4}.}
\label{fig5}
\end{figure}

The electron sublattice of singlet pairs remains 
insulating and preserves spin ordering because a half of electrons of this sublattice is 
in the hybrid orbitals of the same $Mn^{3+}$ ions in which the electrons of triplet pairs 
are present (Fig.\ \ref{fig3}). 

\section{Ferromagnetic fluctuations and the effect of external magnetic field on $T_c$}

The above discussion shows that $T_c$ is the temperature of formation of a 
spin-ordered electron lattice incorporating a spin-polarized electron sublattice of triplet 
pairs. It is evident that this temperature must be preceded by a temperature interval in 
which electrons are still localized in anion orbitals. Though anions polarized by Mn 
ions can be in the 3s2p hybridization state, static overlap of these hybrid orbitals with 
hybrid cation orbitals is absent. Only dynamic overlap due to thermal vibrations of  
ions is possible. It can manifest itself as short-range ferromagnetic fluctuations 
observed experimentally. \cite{16} In this temperature interval, the "foundation" for 
formation of static $1e-\sigma^\ast$ bonds and a spin-polarized sublattice via
a gradual spin polarization of all the electrons of triplet pairs is laid. Apparently, 
application of an external magnetic field, which aligns spins of all unpaired electrons, 
can accelerate the formation of the spin-polarized electron sublattice and thereby 
increase $T_c$. For similar reasons, $T_c$ must also increase with increasing magnetic 
field. \cite{17} 

In terms of the suggested model, the effect of colossal magnetoresistance is the 
consequence of the change of the conductivity mechanisms, i.e., from the thermally 
activated conductivity typical of a solid with preferentially ionic bonding to the 
metallic conductivity caused by a spin-polarized electron sublattice of triplet pairs 
whose formation is facilitated by the external magnetic field.

\section{Conclusions}

The above consideration shows that the origin of a strongly correlated state of 
electrons in manganites responsible for magnetic ordering and metallic conductivity 
can be understood if we represent a strong covalency of chemical bonding between 
manganese and oxygen ions by formation of one-electron $\sigma$ bonds between these 
ions. These bonds are formed by overlapping of empty $d^2sp^3$ orbitals of Mn ions 
with orbitals of anions. Each anion can form two one-electron $\sigma$ bonds with two 
neighboring Mn ions. It is assumed in this case that a $Mn^{3+}$ ion prefers a low-spin 
state and can polarize an anion thereby causing its 3s2p hybridization which is 
accompanied by transition of its singlet electron pair into a triplet state. Both singlet 
and triplet pairs can participate in covalent one-electron bonding, which results in 
different spin ordering of electrons in hybrid orbitals of neighboring cations. The 
correlated state of the electrons forming one-electron $\sigma$ bonds arises due to the 
exchange interaction between electrons when they are simultaneously in hybrid 
orbitals of a cation and in the p orbitals (or hybrid 3s2p orbitals) of anions. This 
exchange interaction leads to both the spin ordering and spatial ordering of 
electrons, which can be represented as formation of a spin-ordered electron lattice.   
The conductivity produced by such an electron lattice depends on the type of spin 
ordering, lattice dimensionality,  and the localization energy, which in turn depends 
on the one-electron $\sigma$ bond energy.  

The model of chemical bonding suggested here can prove useful for explanation of 
high-$T_c$ superconductivity in cuprates. In \cite{18}, an attempt was made to 
represent the bonding between Cu and O ions whose covalency is stronger than in 
manganites through a resonance state of ionic and ordinary two-electron $\sigma$ bonds 
(in analogy with bonding in molecules \cite{12}). In spite of the fact that such 
representation led to the model of additional $\pi$ bonding between Cu and O ions 
which qualitatively explained high-$T_c$ superconductivity, many assumptions of this 
model remained unclear and even controversial. 

The use of the notion of one-electron $\sigma$ bonding to explain superconductivity can 
also clarify the picture of the behavior of correlated electrons in cuprate superconductors. 
With this notion, it becomes clear that, in the $CuO_2$ plane, formation of one-electron 
$\sigma$ bonds is also possible. The anion is also subjected to polarization by cations. 
However, for superconductivity, the anion polarization in a strongly asymmetric crystal 
field normal to the $CuO_2$ plane proves to play a decisive role. It promotes the 3s2p 
hybridization of an anion and forces its triplet electron pair to transfer from two hybrid 
orbitals into one hybrid orbital pointed to a positively charged plane (for instance, 
$Y^{3+}$). In this case, the spin state of this electron pair is transformed from a triplet 
state into a singlet one. These singlet pairs which are in the 3s2p hybrid orbitals of 
anions form a localized electron lattice. Displacement of this lattice, i.e., collective 
displacement of singlet pairs, can occur after $\pi$ overlapping of these 3s2p hybrid 
orbitals with empty hybrid orbitals of $Cu^{3+}$ ions and formation of delocalized $\pi$ 
orbitals. Thus the pairing mechanism in high-$T_c$ superconductors can be the exchange 
interaction between electrons in a hybrid orbital of an anion polarized by a strongly 
asymmetric crystal field. In this model, the mechanism of superconductivity is displacement
of the lattice of electron pairs in delocalized $\pi$ orbitals that belong to the chains of 
$Cu^{3+}$ and $O^{2-}$ ions in the $CuO_2$ plane.

\end{document}